# The role of 2s Bloch wave state excitations on STEM-HAADF intensity in quantitative analysis of alloys

C. Wouters[1], T. Markurt[1], E. Rotunno[2], V. Grillo[2], M. Albrecht[1]

[1]Leibniz-Institut für Kristallzüchtung, Max-Born-Straße 2, 12489 Berlin, Germany

[2]CNR-NANO S3, via Campi 213/A, 41100 Modena, Italy

**Abstract**

**In this work, we emphasize the important contribution of the 2s Bloch wave state to the properties of a STEM electron probe propagating on an atomic column. For a strong enough column potential, the confinement of the 2s state leads to a long-period oscillation of the electron wave function, which is reflected in the resulting STEM-HAADF intensity. We show how this influences STEM composition quantification even at large thicknesses. We found additionally that the excitation of the 2s state affects the intensity of alloys where long-range order phenomena are present, which in turn provides a way to probe the degree of order in alloys.**

Introduction

Over the last decades, Scanning Transmission Electron Microscopy (STEM) has become a well-established technique for providing high-resolution structural and analytical information of solid crystalline materials. Especially the incoherent high angle annular dark field (HAADF) imaging mode, in which strong contrast related to the average atomic number of the probed material is achieved [1, 2], is being used successfully for quantitative chemical composition analysis on the atomic scale. In this imaging mode, a convergent electron beam is focused on the specimen, typically along a low order zone-axis orientation such that electrons propagate along a string of atoms, and electrons scattered to high angles are collected by an annular detector. One of the most interesting results of such zone-axis illumination is the fact that the wave function of the propagating electron beam can be expressed in the form of eigenstates of the Schrödinger equation with the 2D projected structure of the crystal as potential [3]. Because of the periodicity of the crystal potential, these columnar eigenstates can be written as Bloch waves. They are catalogued with analogy to atomic orbitals and are numbered according to a radial and angular quantum number (1s, 2p, 2s, etc.), as illustrated in Fig. 1. For a convergent STEM probe, the most relevant state that is excited is the lowest lying energy eigenstate 1s, of which the amplitude is strongly peaked on the centre of the atomic column. The so-called 1s-state approximation [4] has proven to give a good description of the main features of the wave function inside a crystal and the resulting STEM-HAADF image [5, 6]. However, Anstis and others have shown that this approximation is not always valid and non-1s state contributions can become important for certain incident probe profiles [7] or when the column spacing is not sufficient to exclude excitation of neighbouring columns [8]. Also Rafferty et al. [9] hinted already to an important contribution of the 2s Bloch wave states for heavy columns. We will show indeed, by means of Bloch wave calculations and multislice simulations of isolated atomic columns, that the 2s state becomes bound by the potential for heavy columns and modifies the usual description of propagation in the crystal and the resulting HAADF intensity.

As we will see, the contribution of the 2s Bloch wave state has its consequences for STEM-HAADF analysis in alloy structures. Usually, thick specimens are suited for composition quantification because the 1s channeling of the electron beam, [10] which causes an oscillating variation in beam-atom interaction along the propagation direction, has died out. Therefore, the averaged intensity contrast with respect to a reference material becomes constant and the contrast scales more or less with the square of the average atomic number of the material, without major importance of the atomic arrangement. This assumption has been proven to work, resulting for example in the successful composition quantification in $Al_xGa_{1-x}N$ alloys [11, 12]. What we discover now however, is that for high Z alloys where the 2s excitation comes into play, this is not true anymore. In this case, the dependency of the HAADF intensity contrast on thickness remains for a very large thickness range, which complicates composition quantification. Additionally, we show that also in the large thickness regime the atom arrangement in alloys can have an influence on the HAADF intensity. When imaged in certain zone axis directions, a significant difference in HAADF intensity arises at large thicknesses whether the alloy is ordered or disordered. This provides a method to quantify the amount of long-range order in alloys of known composition on thicker samples, similar to the way the 1s channeling can be exploited to determine 3D atomic configurations on thin samples [13-16]. As ordering phenomena have been observed in different types of materials and it forms an important parameter determining material properties like mechanical, electrical or magnetic behaviour [17-19], this is a very interesting asset of the excitation of the 2s Bloch wave state.

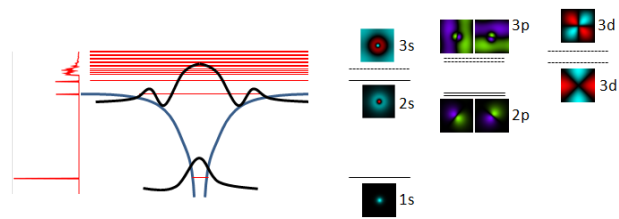

*Figure 1: Right: Representation of the 2D projected potential of an atomic column in a crystal, with bound (localized) and unbound energy eigenstates. Left: Nomenclature of the columnar Bloch wave eigenstates with their spatial amplitude distribution.*

**Methods**

Bloch wave calculations and their spectral description are performed using our custom software B_WISE. [20] Since Bloch wave algorithms only work for plane wave illumination conditions, our software samples a number of points within the probe: the spectra reported in this work have been calculated sampling a STEM probe obtained for a semi-convergence angle of 9 mrad and high voltage of 300 kV into 2445 individual points. For each one of these points we performed a Bloch wave calculation following the original algorithm proposed by Metherell [21] and then summed the resulting Bloch coefficients together taking into account the appropriate aberration phase.

STEM-HAADF image simulations are performed using a parallelized multislice approach [2, 22] with the frozen phonon approximation to account for thermal vibrations of the atoms. Again a semi-convergence angle of 9 mrad and high voltage of 300 kV are chosen, which also matches typical experimental conditions. All simulations are performed by averaging over 40 frozen phonon configurations and the used supercells extend minimally 3x3 nm^2 perpendicular to the beam direction (to ensure a small enough sampling in k-space) and are constructed with periodic boundary conditions in the x-, y-directions. The interatomic spacing along the column is taken to be 3 Å, 3.212 Å, 3.189 Å and 3.753 Å for the supercells of the isolated column, the InGaO3 (monoclinic), In0.5Ga0.5N (wurtzite) and AuCu3 (cubic) lattice, respectively. Along the beam direction, the supercell thickness extends over approximately 30 nm and is repeated multiple times to probe the final thickness. In this way, statistical incorporation of the different constituent atoms along the columns in case of a disordered lattice, creates a sample that can be assumed to be a random configuration without producing any periodicity effects. Although we are aware of the importance of static displacements in analysis of alloys [23, 12], we purposely remove them from our simulations to single out the effect of the 2s excitation on the HAADF intensity.

**Excitation of 2s state**

Results of Bloch wave calculations and multislice simulations of a convergent STEM electron probe propagating on isolated Ga (Z=31), Zr (Z=40) and In (Z=49) atomic columns are shown in Fig. 2, to demonstrate the changes with increasing Z value. In the left panel, the Bloch wave excitation amplitude as a function of transverse energy ($E_T$) is compared for the three cases. The transverse energy is defined as the difference between the z component of the electron wave kinetic energy in the sample and in the vacuum, and is thus proportional to the square of the wave vector $k_z$ along the propagation direction ($E_T \sim k_z^2 - k_0^2$). As illustrated in Fig. 1, a distinction can be made between bound and unbound states. The former are localized by the potential on the atomic columns and form sharp lines in the energy spectrum. The latter are delocalized plane waves and are contained in the excitation energy continuum at lower energies. The amount of bound states and their transverse energy depends on the depth of the potential. For the Ga column, only the 1s state is bound by the potential. As the average atomic number of the column increases, the potential becomes stronger and the next lowest energy eigenstates will start to get confined. For the Zr and especially the In column, the potential is deep enough to confine also the 2p and 2s states. We also see that the transverse energy of the states increases drastically for heavier atoms due to a stronger localization.

The influence of this difference in Bloch wave excitations with Z on the behaviour of the total wave function, is apparent in the right panel in Fig. 2, which shows a cut through the intensity (=$|\psi|^2$) of the electron wave function as a function of thickness (on abscises) and spatial coordinate (on ordinates). The electron probe is placed exactly at the centre of the atomic column. Beating of the 1s state with the unbound states produces the wave function oscillations at small thicknesses in each case. As the atomic number of the column increases, the beating frequency increases (see Table I) and hence the extinction length decreases. In case of the Ga column, the oscillations die out due to dephasing after a thickness of approximately 30 nm, i.e. at this point the electron probe delocalizes from the atomic column and spreads into the surroundings. For the Zr and In columns, similar behaviour is observed for the 1s state with a smaller extinction length, but an additional intensity oscillation is produced by the beating of the bound 2s state with the unbound states. Since the excitation energies of the 2s and unbound states are closer to each other, a larger oscillation wavelength results for this interaction, as summarized in Table I. Apart from a wave function maximum located at the column centre, the 2s state has a second radially symmetric maximum at a radius of about 0.9 Å, as apparent in Fig. 1. The 2p state, which also becomes a bound eigenstate for the heavier columns, doesn't significantly contribute to the HAADF signal because it has a node in the centre, directly above the atomic column, and therefore it will be neglected throughout the rest of the discussion.

These results show clearly how for a fixed atomic spacing there is a threshold in Z where the 2s Bloch wave state starts to get bound by the columnar potential and how this produces a second intensity oscillation, of longer wavelength, that persists to much larger thicknesses. Of course, this threshold depends on additional parameters such as the spacing between the atoms and the initial beam characteristics.

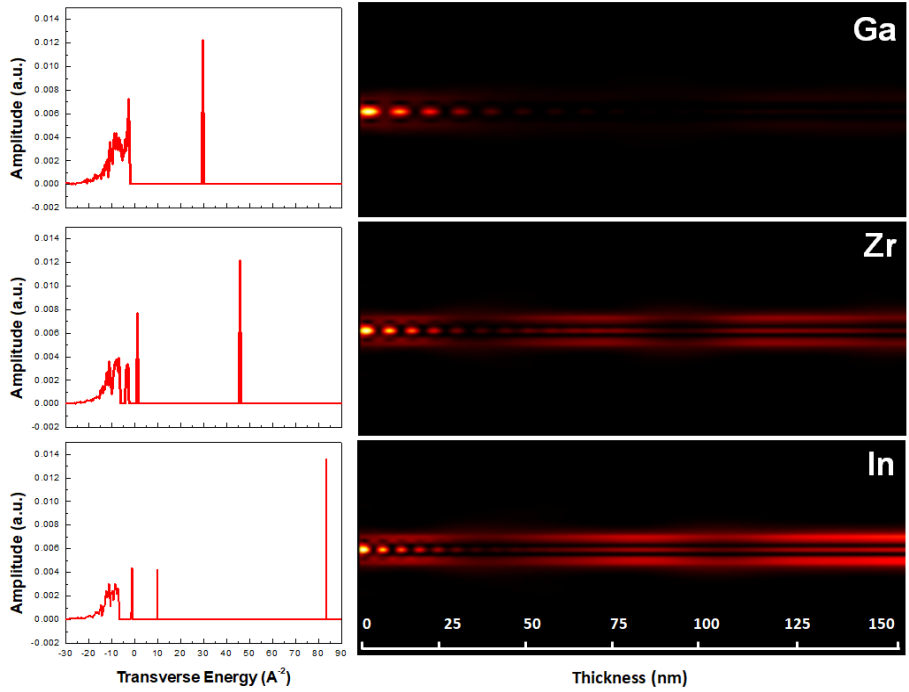

*Figure 2: (Left) Bloch wave excitation spectrum as a function of transverse energy ($E_T$) for a STEM electron probe propagating on an isolated Ga (Z=31), Zr (Z=40) and In (Z=49) atomic column. (Right) Cut through the electron wave function intensity as a function of thickness.*

|  | Ga | Zr | In | $Ga_{0.5}In_{0.5}$ |
|---|---|---|---|---|
| **1s $E_T$ (Å$^{-2}$)** | 28.5 | 45.7 | 83.2 | 62.8 |
| **2s $E_T$ (Å$^{-2}$)** | - | -3.8 | -1 | -2.8 |
| **Unbound States** (UB) $<E_T>$ (Å$^{-2}$) | -8 | -11 | -11 | -11 |
| **1s+UB** Beating wavelength (nm) | 10.8 | 7.2 | 4.29 | 5.5 |
| **2s+UB** Beating wavelength (nm) | - | 56.5 | 40.6 | 50.2 |

*Table 1: Bloch wave properties of a STEM electron probe propagating on different types of isolated atomic columns.*

**Implications for quantitative HAADF imaging**

Let us investigate how the excitation of the bound 2s Bloch wave state is influencing the HAADF intensity as a function of sample thickness. Multislice STEM simulations are performed for two series of isolated atomic columns with increasing average atomic number to study the effect of the on-set of the bound 2s eigenstate. In the first series, the columns consist of one type of element with Z increasing from 22 to 49. In the second series, we start from a pure Ga column and the average atomic number along the column is increased by randomly mixing in In atoms with composition ratios ranging from 0:1. The on-column intensity output recorded by the simulated HAADF detector is plotted as a function of sample thickness in Fig. 3(a) and (b). For both series, the small intensity oscillations at low thicknesses (<40 nm) are caused by the channeling effect that induces the electrons in the probe to be periodically focused by the attractive periodic potential of the atoms in the column. This is the classical particle picture of the wave function oscillations that were earlier described in the Bloch wave formalism as caused by the beating interaction of the bound 1s state with the unbound states. For larger thicknesses, a strong change in the behaviour of the intensity can be observed once the average atomic number along the column exceeds a certain threshold. For the low Z columns the intensity keeps on increasing at a more or less constant rate, while for the high Z columns a sudden increase in intensity takes place. This sudden increase can be attributed to the excitation of the bound 2s Bloch wave state, for which, as illustrated before, the same threshold around $Z \approx 36$ was observed for this specific atomic spacing. The onset of this feature takes places at a thickness of around 40-60 nm, which corresponds approximately to the beginning of the second beating period of the 2s-unbound interaction (see Table I). This interaction causes long-wavelength oscillations in the intensity to persist more strongly as the average atomic number of the column increases. The fact that the same general trend is observed in both the single-atom type series and the increasing composition series consisting of a mix of two elements indicates that the onset of confinement of the 2s state is mostly determined by the average atomic density along the column. However, just like the 1s oscillations, we expect that the 2s oscillation causes some dependency of the HAADF intensity on the local distribution of atomic number density along the depth of the column.

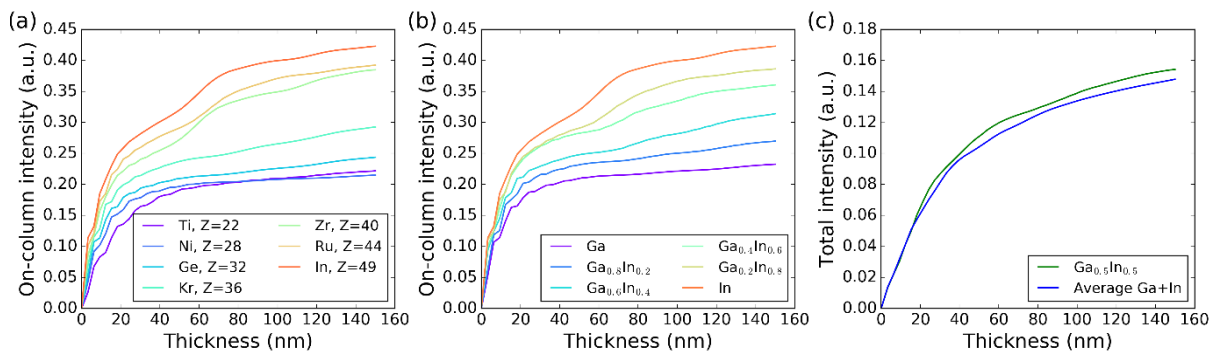

*Figure 3: Simulated on-column STEM-HAADF intensity of isolated atomic columns for (a) columns containing a single type of element with 22<Z<49 and (b) columns containing a mix of Ga and In atoms in different composition ratio's in a random configuration with 31<$Z_{avg}$<49. (c) Comparison of total intensity of a column containing an equal mix of Ga and In atoms versus the average total intensity of a pure Ga and In column.*

Since heavy columns are strongly affected by the 2s Bloch wave state excitation, let us see what the implications are for STEM-HAADF intensity analysis in high Z alloys. As the STEM-HAADF intensity scales roughly with average atomic number, it is used to quantify composition of alloys by comparing the average intensity in the experimental images to that of simulated images. Typically, the intensity ratio with respect to a reference material is calculated, since this value becomes constant for large thicknesses and knowledge of the exact thickness of the specimen is not necessary for comparison. This is nicely illustrated in Ref. [11] for AlxGa1-xN alloys of different compositions. In Fig. 4, we compare similar intensity ratios in randomly configured monoclinic (GaxAl1-x)2O3 and (InxGa1-x)2O3 alloys. In the case of (GaxAl1-x)2O3, the intensity ratio of (Ga0.5Al0.5)2O3 and Ga2O3 to Al2O3 becomes constant for thicknesses higher than approximately 100 nm. The strong oscillations at small thicknesses are caused by the differences in channeling behaviour. However, if we compare to (InxGa1-x)2O3 alloys, the desired behaviour of a constant intensity ratio is not present anymore. The thickness dependency of the intensity contrast remains up to thicknesses of 200 nm. After the channeling oscillations, the contrast doesn't saturate to a constant value, but due to the long-wavelength 2s oscillations that come into play for these heavier materials, strong contrast variations remain at large thicknesses. The reason for the steady decrease of the contrast at thicknesses >100 nm is currently under investigation. Due to this behaviour, composition quantification in such heavier systems becomes a lot more difficult since specimen thickness has to be a well-known parameter to connect the intensity ratio to a composition.

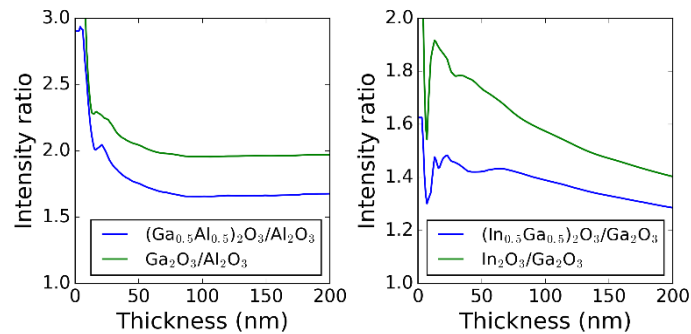

*Figure 4: (Ga$_x$Al$_{1-x}$)$_2$O$_3$/ Al$_2$O$_3$ and (In$_x$Ga$_{1-x}$)$_2$O$_3$/ Ga$_2$O$_3$ intensity ratio's for x=0.5 and x=1 plotted as a function of specimen thickness.*

**Ordered vs. disordered alloys**

Another important and interesting consequence of the extra 'jump' in HAADF intensity that appears for heavy columns due to the 2s state excitation, is observed in ordered alloys. Comparing ordered and disordered alloy structures, i.e. distribution of each of the constituent atoms on a distinct sublattice versus random distribution of both constituent atoms on all possible lattice sites, a significant difference in HAADF intensity arises at large thicknesses. InGaO3 ((InxGa1-x)2O3 with x=0.5), In0.5Ga0.5N and AuCu3 alloys are considered, to show the effect in three different lattice symmetries: monoclinic, wurtzite and cubic (fcc), respectively. In the monoclinic lattice of InGaO3, an equal amount of two types of lattice sites exist for the cations, which differ in their coordination to the oxygen atoms (tetrahedral vs. octahedral). Due to the strong preference of the indium atoms for an octahedral environment [24], a sub-lattice ordering is created with all indium atoms occupying the octahedral positions and all gallium atoms the tetrahedral positions. In AuCu3, a phase transition to an ordered state takes places below a certain transition temperature (< 390°C), where the Au atoms prefer to be surrounded by Cu atoms as nearest neighbour and they are positioned exclusively on the corners of the face-centered cubic (fcc) unit cell [25]. In contrast to these two intrinsic ordering phenomena, artificial ordering was assumed in the case of In0.5Ga0.5N. Ordered and disordered supercells are constructed as described in the Methods section and the projected supercell perpendicular to the beam direction is visualized for the case of InGaO3 in Fig. 5(a) and (b). One unit cell (dashed shape) and the scanning area for the simulation (filled area) are indicated. In the ordered structure, the electron beam 'sees' only columns which consist of one type of atom, while in the disordered cell, each cation column consists of a random stoichiometric distribution of the two constituent atoms. This is true for each of the considered materials for the chosen beam directions. For each of the alloys, ordered and disordered, the average intensity over one unit cell (an approximate unit cell in the case of the monoclinic structure, since we can only scan rectangular cells) is extracted as a function of thickness and plotted in Fig. 5(c). The same trend is observed everywhere: for thicknesses > 40 nm, HAADF intensities of ordered and disordered structures start to diverge, with the disordered lattice always having the higher intensity. The percentage difference between the ordered and disordered intensities at a thickness of 100 nm ranges between 7-12.5% for the three systems. The thickness on-set of the divergence corresponds exactly to the characteristic thickness where the low frequency intensity oscillation caused by the excitation of the 2s state starts to dominate.

To explain this remarkable phenomenon, we consider again some isolated column simulations. In Fig. 3(c), the total HAADF intensity of a randomly configured In0.5Ga0.5 column is compared to the average total intensity of a pure Ga and In column. While in the channeling regime the intensities are still as good as equal, for larger thicknesses the randomly configured In0.5Ga0.5 column consistently has the higher intensity. This can be understood as follows: for the pure Ga column the 2s state is not contributing, but for the mixed In0.5Ga0.5 column and the pure In column it is (see Table 1). The lack of the 2s intensity oscillation for the Ga column results in a lower intensity when averaged with an In column, compared to the mixed column. Now let us assume that for the InGaO3 lattice, we can approximate the total intensity as the sum of the intensities of the isolated cation columns. Since the atomic spacing along the columns in the monoclinic lattice is close to the 3Å considered in the isolated columns, the general outcome of the results can be transferred. This means that in the ordered lattice, the 2s state is excited for only half of the cation columns (In columns), while in the disordered structure, the 2s state is excited for all cation columns (In0.5Ga0.5). Following the result of Fig. 3(c), a higher intensity should indeed be expected for the disordered lattice. The same explanation accounts for the In0.5Ga0.5N and AuCu3 structures due to the mixture of one low Z (i.e. no 2s excitation) and one high Z (i.e. 2s excitation) element. Of course, in the complete lattice structures, there are more factors playing a role like cross-correlation between neighbouring columns, symmetry of the lattice, etc. Hence, every material or orientation is a very specific case and needs to be treated individually.

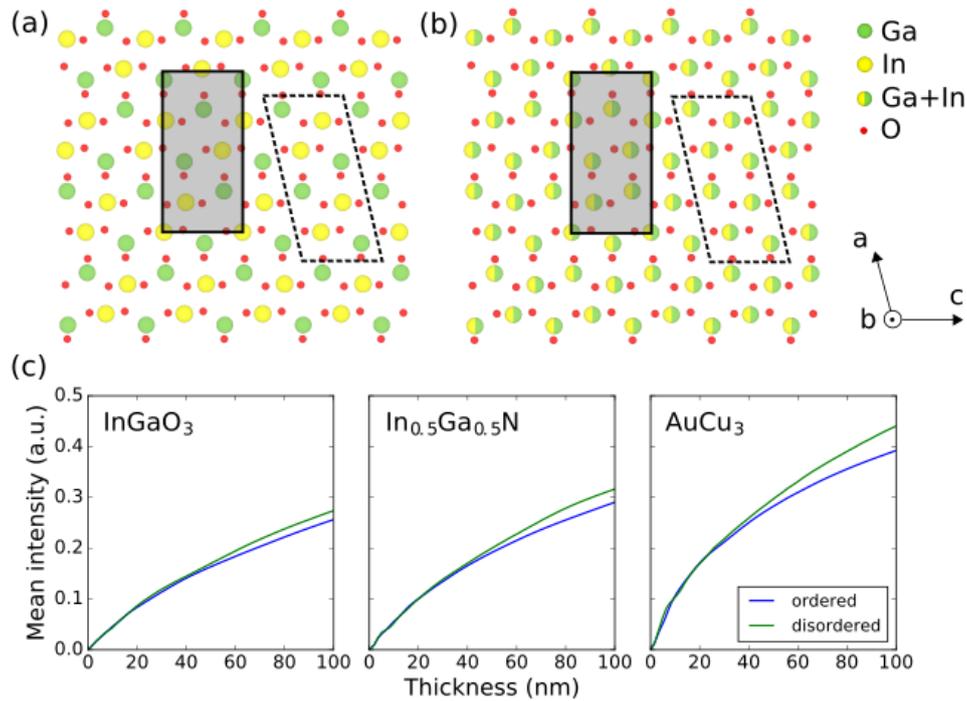

*Figure 5: a&b) Schematic showing the geometry of the supercells (not full size) used for multislice STEM-HAADF simulations of ordered, consisting of pure Ga and In columns, and disordered, consisting of mixed In+Ga columns, InGaO$_3$ structures projected perpendicular to the beam direction (b-axis). c) Mean intensity as a function of thickness for an ordered vs. disordered unit cell of InGaO$_3$, In$_{0.5}$Ga$_{0.5}$N and AuCu$_3$.*

As a consequence of this order-disorder intensity difference, composition quantification could become complicated when the ordering in the alloy system under study is unknown. However, when the composition is known, this phenomenon could possibly be exploited to our advantage. The dependency of the HAADF intensity on the ordering in the system could be used to estimate the amount of order in materials by comparing experimental STEM HAADF images to simulations. To see if this holds up, additional supercells were created for InGaO3, In0.5Ga0.5N and AuCu3 structures with varying degree of order. To describe the amount of order, we introduce a long-range order parameter S, as defined by Cowley et al. [26], which quantifies the amount of atoms that are occupying their 'correct' position in the lattice. S=0 means a completely random distribution of atoms; S=1 means a perfectly ordered crystal. HAADF intensities are determined at a thickness of 100 nm and their dependency to S are shown in Fig. 6. For InGaO3 and In0.5Ga0.5N, a monotonic decrease in intensity is found as the order parameter increases. In the case of AuCu3, the intensity increases slightly as some small amount of order is introduced, but for S>0.5 a significant and monotonic decrease of intensity is observed. A parabolic curve gives a good fit to all data sets.

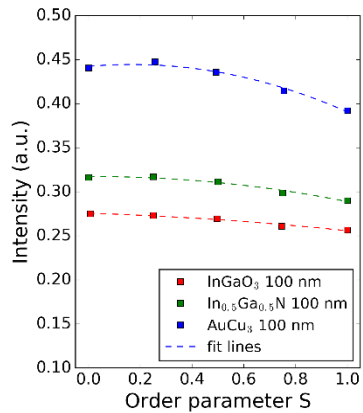

*Figure 6: Simulated STEM-HAADF intensity at a thickness of 100 nm as a function of order parameter for InGaO$_3$ and AuCu$_3$ alloys with parabolic fit lines to the data.*

We note that all results of the multislice simulations use an inner-acceptance angle of the HAADF detector of 35 mrad, because this corresponds to our typical experimental conditions. However, we analysed the intensities also for other scattering angles and found the same results (see SI Fig S1&S2).

**Conclusion**

In conclusion, we illustrated the importance of the excitation of the bound 2s Bloch wave eigenstate in a STEM electron probe propagating on an atomic column. Just like the 1s excitation, it produces an oscillation - in this case of longer wavelength - of the electron wave function due to the interference with the unbound Bloch wave states. This oscillation is strongly reflected in the STEM HAADF intensity where it persists up to thicknesses larger than 100 nm. As a result, intensity contrast in heavy alloys is strongly modulated up to large thicknesses which complicates composition quantification. We highlighted another important consequence for alloys that consist of a mixture of elements of relatively low and high Z. When the alloy is ordered and imaged in a zone-axis direction where all columns consist of the same atom, its average intensity is systematically lower than when the alloy is disordered, starting from thicknesses < 40 nm. It was shown how this dependency of the STEM HAADF intensity on the order parameter provides a method to estimate the degree of long-range order in these types of alloy systems.

**Acknowledgements**

This work was performed in the framework of GraFOx, a Leibniz-Science Campus partially funded by the Leibniz Association. The research leading to these results has received funding from the European Union´s Horizon 2020 Research and Innovation programme under Grant Agreement No 766970 QSORT (H2020-FETOPEN-1-2016-2017).

**References**

[1] Pennycook S. J. and Boatner L.A. (1988), Nature **336**: 565–567.

[2] Kirkland E. J., R. F. Loane, and J. Silcox, Ultramicroscopy **23**, 77 (1987).

[3] B. F. Buxton, J. E. Loveluck, and J. W. Steeds, Philos. Mag. A **38**, 259 (1978).

[4] P. Geuens and D. Van Dyck, The S-State model: a work horse for HRTEM, Ultramicroscopy **93** (2002) 179-198.

[5] Pennycook S. J. and Jesson D.E., Phys. Rev. Lett. **64**, 8 (1990) 938–941.

[6] Peng Y., Nellist P.D. and S.J. Pennycook, J. Electron Microsc. **53** (3), 257–266 (2004).

[7] G.R. Anstis, Microsc. Microanal. **10**, 4–8 (2004).

[8] G.R. Anstis, D.Q. Cai, D.J.H. Cockayne, Ultramicroscopy **94** (2003) 309–327.

[9] B. Rafferty, P.D. Nellist and S. J. Pennycook, Journal of Electron Microscopy **50**(3): 227–233 (2001).

[10] Howie A., Philos. Mag. **14** (1967) 223.

[11] Rosenauer A. et al., Ultramicroscopy **109** (2009) 1171–1182.

[12] M. Schowalter et al., Microsc. Microanal. **20**, 1463–1470, 2014.

[13] E. Rotunno et al., Ultramicroscopy **146** (2014) 62–70.

[14] Voyles P.M., Grazul J.L. and Muller D.A., Ultramicroscopy **96** (2003) 251–273.

[15] Hwang J. et al., Phys. Rev. Lett. **111**, 266101 (2013).

[16] Ishikawa R. et al., Nano Lett. **14**, 1903–1908 (2014).

[17] Hauser A.J. et al., Phys. Rev. B **83**, 014407 (2011).

[18] B. D. Esser et al., Phys. Rev. Letters **117**, 176101 (2016).

[19] J.B. Cohen, Journal of Materials Science **4** (1969) 1012-1022.

[20] E. Rotunno, A.H. Tavabi, E. Yucelen, S. Frabboni, R.E. Dunin Borkowski, E. Karimi, B.J. McMorran, and V. Grillo, Electron-Beam Shaping in the Transmission Electron Microscope: Control of Electron-Beam Propagation Along Atomic Columns, Phys. Rev. Applied 11, 04407 (2019)

[21] A. J. F. Metherell in Electron Microscopy in Materials Science II (1975) 397–552 Eds. U. Valdrè and E. Ruedl, CEC, Brussels.

[22] Kirkland E. J., Advanced Computing in Electron Microscopy (Springer, New York, 2010) 2nd ed.

[23] Grillo V., Carlino E. and Glas F., Phys. Rev. B **77**, 054103 (2008).

[24] Maccioni M.B. et al, Appl. Phys. Express **8**, 021102 (2015).

[25] Claeson T. and Boyce J.B., Phys. Rev. B **29**, 4 (1982).

[26] Cowley J.M., Phys. Rev. **77**, 5 (1950).

---------------------------

**Supplementary information**

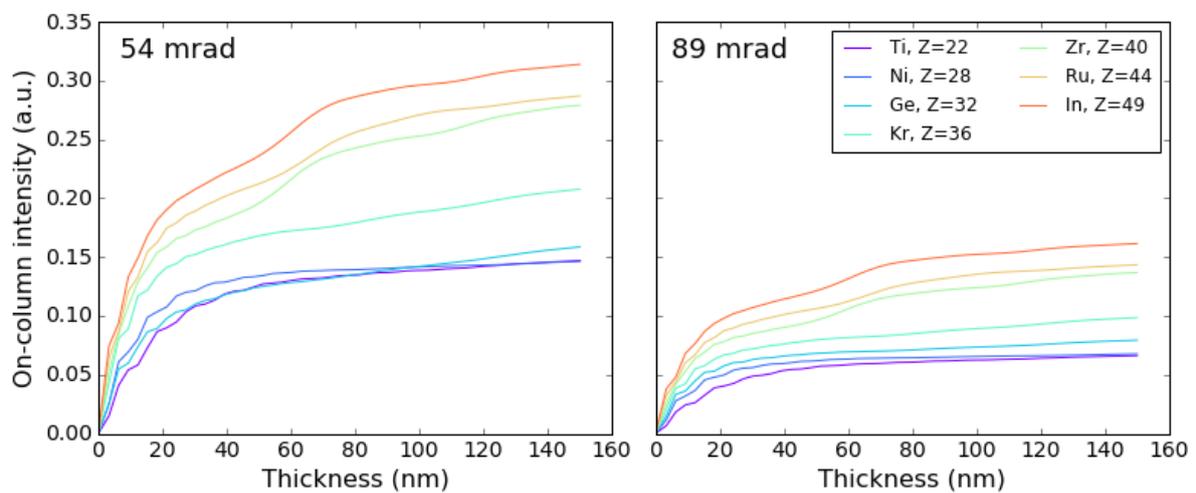

*Figure S1: Simulated on-column STEM-HAADF intensity of single-type isolated atomic columns with 22≤Z≤49, for two different inner-acceptance angles of the detector.*

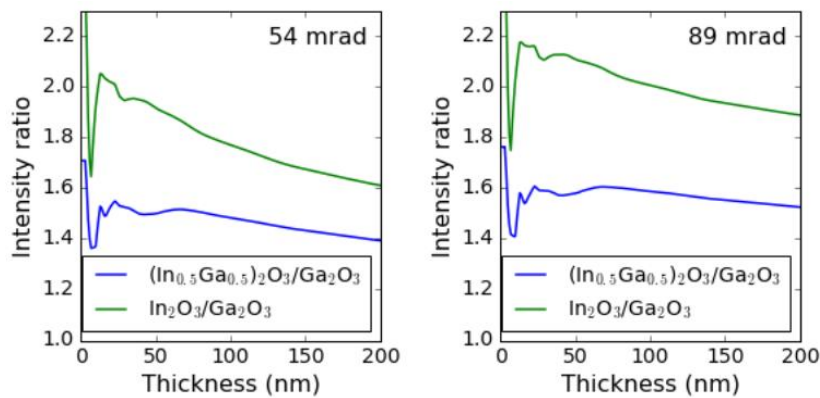

Figure S2: InxGa1-x)2O3/ Ga2O3 intensity ratio's for x=0.5 and x=1 plotted as a function of specimen thickness, for two different inner-acceptance angles of the detector..

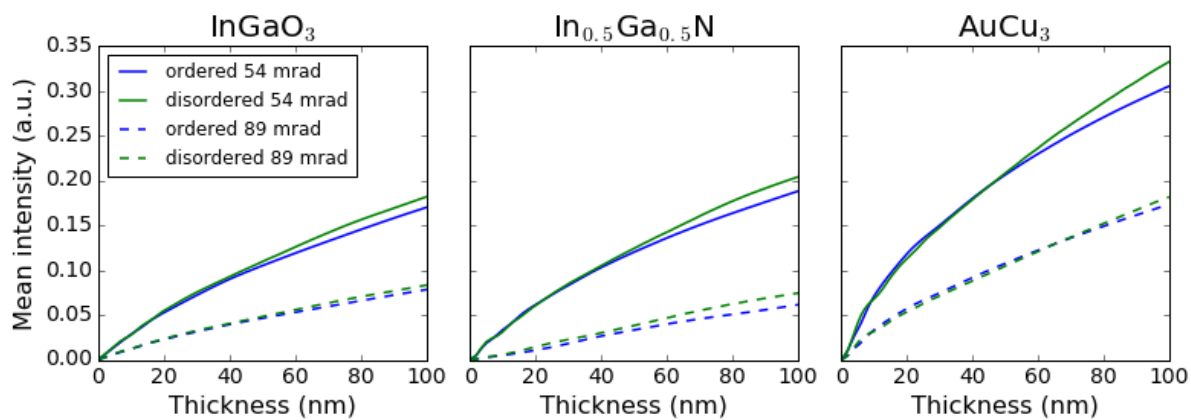

Figure S3: Mean intensity as a function of thickness for an ordered vs. disordered unit cell of $InGaO_3$, $In_{0.5}Ga_{0.5}N$ and $AuCu_3$, for two different inner-acceptance angles of the detector.